\documentclass[conference]{IEEEtran}
\usepackage{cite}
\usepackage{soul}
\usepackage{amsmath,amssymb,amsfonts}
\usepackage{algorithmic}
\usepackage{graphicx}
\usepackage{textcomp}
\usepackage{xcolor}
\usepackage[center]{caption}
\usepackage{subfig}
\usepackage{multirow}
\usepackage{graphicx}
\usepackage{pifont}

\def\BibTeX{{\rm B\kern-.05em{\sc i\kern-.025em b}\kern-.08em
    T\kern-.1667em\lower.7ex\hbox{E}\kern-.125emX}}
\begin{document}

\title{
The Lynchpin of In-Memory Computing: \\
A Benchmarking Framework for Vector-Matrix Multiplication in RRAMs

}
\makeatletter 
\newcommand{\linebreakand}{%
  \end{@IEEEauthorhalign}
  \hfill\mbox{}\par
  \mbox{}\hfill\begin{@IEEEauthorhalign}
}
\makeatother 
\author{

\IEEEauthorblockN{Md Tawsif Rahman Chowdhury}
\IEEEauthorblockA{\textit{Electrical and Computer Engineering} \\
\textit{Wayne State University}\\
Detroit, USA\\
mtawsifrc@wayne.edu}
\and
\IEEEauthorblockN{Huynh Quang Nguyen Vo}
\IEEEauthorblockA{\textit{Industrial Engineering and Management} \\
\textit{Oklahoma State University}\\
Stillwater, USA\\
lucius.vo@okstate.edu}
\and
\IEEEauthorblockN{Paritosh Ramanan}
\IEEEauthorblockA{\textit{Industrial Engineering and Management} \\
\textit{Oklahoma State University}\\
Stillwater, USA\\
paritosh.ramanan@okstate.edu}
\and
\hspace{3cm}
\IEEEauthorblockN{Murat Yildirim} 
\IEEEauthorblockA{\textit{\hspace{3cm}Industrial and Systems Engineering} \\
\textit{\hspace{3cm}Wayne State University}\\
\hspace{3cm}Detroit, USA\\
\hspace{3cm}murat@wayne.edu}
\and
\IEEEauthorblockN{Gozde Tutuncuoglu}
\IEEEauthorblockA{\textit{
Electrical and Computer Engineering} \\
\textit{Wayne State University}\\
Detroit, USA\\
gozde@wayne.edu}
}
\maketitle

\begin{abstract}

The Von Neumann bottleneck, a fundamental challenge in conventional computer architecture, arises from the inability to execute fetch and data operations simultaneously due to a shared bus linking processing and memory units. This bottleneck significantly limits system performance, increases energy consumption, and exacerbates computational complexity. Emerging technologies such as Resistive Random Access Memories (RRAMs), leveraging crossbar arrays, offer promising alternatives for addressing the demands of data-intensive computational tasks through in-memory computing of analog vector-matrix multiplication (VMM) operations. However, the propagation of errors due to device and circuit-level imperfections remains a significant challenge. In this study, we introduce MELISO (In-Memory Linear Solver), a comprehensive end-to-end VMM benchmarking framework tailored for RRAM-based systems. MELISO evaluates the error propagation in VMM operations, analyzing the impact of RRAM device metrics on error magnitude and distribution. This paper introduces the MELISO framework and demonstrates its utility in characterizing and mitigating VMM error propagation using state-of-the-art RRAM device metrics.

\end{abstract}


\section{Introduction}

The proliferating computation complexities and the associated energy consumption patterns are becoming a pressing concern in today's digital era. Projections indicate that by 2040, the cumulative energy consumption for computer operations could soar to an astonishing $10^{27}$ Joules, surpassing the anticipated capacity for energy production \cite{bx1}. Conventional computing systems, rooted in the Von Neumann architecture, encounter fundamental obstacles in keeping pace with this exponential growth. The \textit{Von Neumann bottleneck} emerges as a pivotal constraint, wherein the simultaneous execution of fetch and data operations grinds to a halt due to the bottleneck imposed by a shared bus connecting the processing and memory units \cite{b1,b2,b3}. Consequently, system performance is throttled, exacerbating energy consumption and imposing limitations on computational complexity. A recent study investigating Google server workloads reports that 62.7\% of the total energy was allocated to data movements rather than computations in these systems \cite{bx2}. This inefficiency not only exacerbates energy consumption but also imposes a significant computational overhead, impeding the realization of computational tasks with higher complexity.



Emerging non-volatile memory technologies, such as resistive random access memories (RRAMs) based crossbar arrays, provide a versatile solution to the challenges of data-intensive, large-scale computational tasks \cite{b4,b6}. By co-locating memory and computation nodes (i.e. performing in-memory computing), RRAM computing eliminates the need for resource-intensive communication tasks, thereby fundamentally improving computational complexity and energy use.

\begin{figure*}[!htb]
    \centering
    {\label{fig:figure1}\includegraphics[width=.98\textwidth,keepaspectratio]{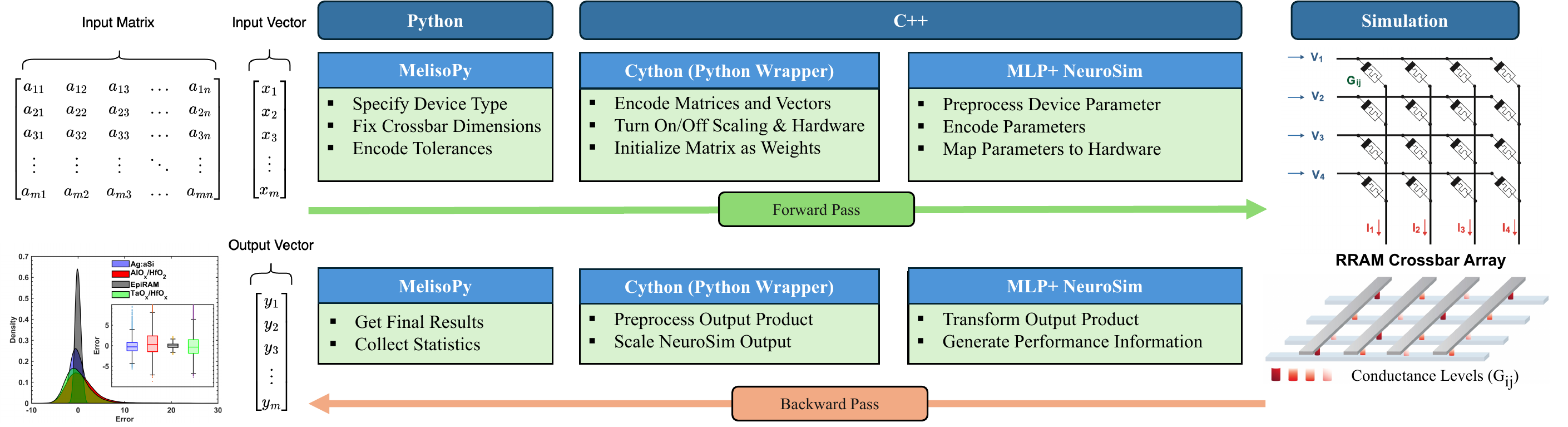}}\\
    \caption{Overview of MELISO: An end-to-end VMM benchmarking framework}
    \label{fig1}
\vspace{-0mm}
\end{figure*}
In a conventional RRAM crossbar array, each word line is connected to the bit line through an RRAM device. This architecture is used to perform vector-matrix multiplication (VMM) operations, the foundational computing block, and {lynchpin} for a myriad of modern computing tasks. Facilitating an analog VMM, the output current at each column, $I_{ij}$, is computed as the sum of the products of the input voltages and the corresponding conductance values of each row, expressed mathematically as $I_{ij}=\sum{V_i \bullet G_{ij}}$. This mechanism underscores the computational efficiency inherent in the RRAM crossbar design for implementing VMM operations, with improved energy efficiency and reduced latency metrics \cite{b7,b8,b9,b10,b11,b12,b13,b14,b15}. Moreover, RRAM reduces the effective computational complexity of dense VMM operations from $O(n^2)$ to $O(1)$. The efficiency gains achieved in VMM operations can serve as the lynchpin and enable cascading benefits across a spectrum of computing tasks. 
These tasks extend beyond conventional neural network computations, providing a highly configurable computational platform for conducting any linear computation tasks, such as solving linear algebra and optimization problems that demand extensive matrix-vector multiplications and linear equations. The combination of low computational complexity, low latency, and energy-efficient operations provides an ideal computational setting for managing large volumes of data that are typical to these applications \cite{ut2}.

It should be noted that a parallel compounding phenomenon also unfolds concerning the propagation of errors within VMM operations due to issues related to device non-idealities (e.g., low-precision RRAMs, and cycle-to-cycle (C-to-C) variations) and the challenges of CMOS integration.  Depending on the algorithmic application, this variability can constitute either a challenge or an asset. For algorithms with strict precision requirements, it becomes an important challenge to devise methods to mitigate this variability and its impact on algorithm performance. For algorithms that require sampling such as Markov Chain Monte Carlo used in Bayesian learning schemes \cite{ut1}, these variabilities can be leveraged as realizations of sampled uncertainties, further reducing the computational burden. It is crucial to understand the relationship between physical device parameters and the algorithm requirements to achieve a holistic algorithm-device co-design framework.


\begin{table}[h]
\caption{State-of-the-Art Device Metrics }
\begin{center}
\small
\begin{tabular}{c|c|c|c|c}
\hline
\textbf{Device Type} & 
\textbf{\textit{Ag:a-Si}}& \textbf{\textit{TaO\textsubscript{x}/HfO\textsubscript{x}}}& \textbf{\textit{AlO\textsubscript{x}/HfO\textsubscript{2}}}&
\textbf{\textit{EpiRAM}}\\
\hline 
CS  & 97 & 128 & 40 & 64 \\
\hline
Non-linearity & 2.4/-4.88 & 0.04/-0.63 & 1.94/-0.61 & 0.5/-0.5 \\
\hline
R\textsubscript{ON} ($\Omega$) & 26M & 100K & 16.9K & 81K \\
\hline
MW & 12.5 & 10 & 4.43 & 50.2 \\
\hline
C-to-C ($\%$) & 3.5 & 3.7 & 5 & 2 \\
\hline
\multicolumn{5}{l}{$^\mathrm{*}$\textit{Ag:a-Si}\cite{b22}, \textit{TaO\textsubscript{x}/HfO\textsubscript{x}}\cite{b23}, \textit{AlO\textsubscript{x}/HfO\textsubscript{2}}\cite{b24}, \textit{EpiRAM}\cite{b25}} \\
\multicolumn{5}{c}{(CS: Conductance States, MW: Memory Window)}
\end{tabular}
\label{tab1}
\end{center}
\end{table}

Extensive literature exists on benchmarking frameworks designed to evaluate RRAM device performance and integration with CMOS peripheral circuitry for a variety of computational tasks such as image classification with fully connected and convolutional neural networks \cite{b16,b17,b18}, as well as dot product engines \cite{b19}. As RRAMs critically suffer from device non-idealities, most prominently non-linear conductance tuning, C-to-C, and device-to-device variations, benchmarking frameworks need to incorporate these key issues \cite{b20,b21}. Functioning as a circuit-level macro model and benchmarking tool, NeuroSim+ serves as a prominent example in this regard, designed for evaluating neuro-inspired architectures \cite{bx3,bx4}. This framework quantifies essential circuit-level performance metrics (e.g., chip area, latency) by integrating modifications at the device output, circuit, and algorithm levels.


Realization of the computational benefits of RRAM devices necessitates a concerted effort toward the identification, modeling, and characterization of the propagation of errors within VMM operations. This characterization paves the way for a new generation of hardware and algorithmic methods to contain or harness these error terms to ensure high levels of computational accuracy. In response to this need, this paper presents a benchmarking framework designed to provide a comprehensive analysis of the error propagation in VMM operations for different RRAM devices and device properties. 
The contributions of this work can be summarized as follows:

\begin{itemize}
\item We develop an end-to-end VMM benchmarking framework for RRAMs, called MELISO - \textit{In-\underline{Me}mory \underline{Li}near \underline{So}lver}. MELISO builds on NeuroSim+ \cite{bx3,bx4} to provide capabilities for compute-in-memory VMM, a foundational mathematical operation that lies at the core of every computing task.
\item We test and analyze the impact of RRAM device parameters and chemistries on the error terms observed in VMM tasks. We implement a comprehensive benchmarking study across a statistically significant population of RRAM devices and VMM operations. We analyze patterns in error distributions with respect to different device chemistries, and device parameters such as C-to-C variability and nonlinearities.
\item We analyze and identify the parametric distributions that best represent the errors in RRAM VMM operations to guide the error assumptions of the subsequent algorithm design efforts that aim to mitigate or leverage error propagation for in-memory computing applications. 
\end{itemize}

The rest of the paper proceeds as follows. Section II introduces the methodology used for the realization of the benchmarking framework. An extensive set of results are demonstrated in Section III. Section IV concludes the paper by elucidating the contributions and future research outlook.

\begin{figure} [h]
    \centering
    \subfloat[]{\label{fig2a}\includegraphics[width=0.49\linewidth]{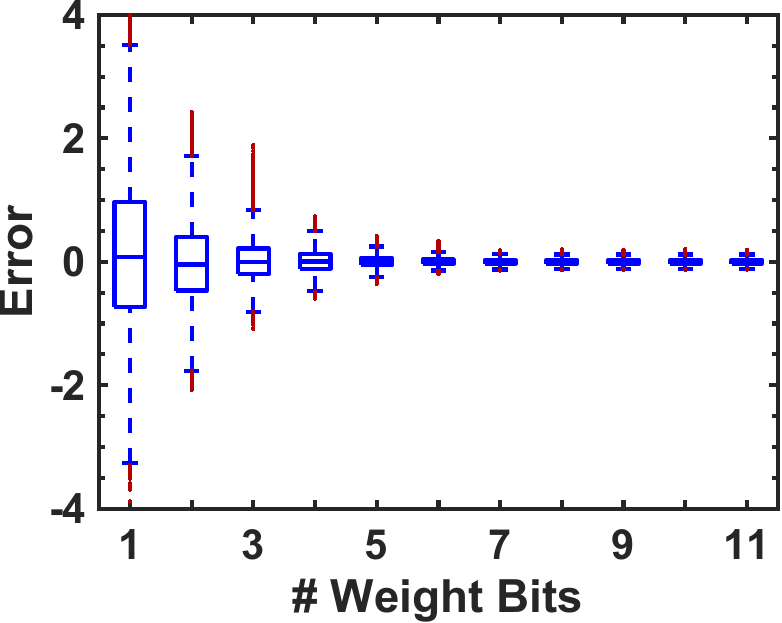}}\hfill
    \subfloat[]{\label{fig2b}\includegraphics[width=0.49\linewidth]{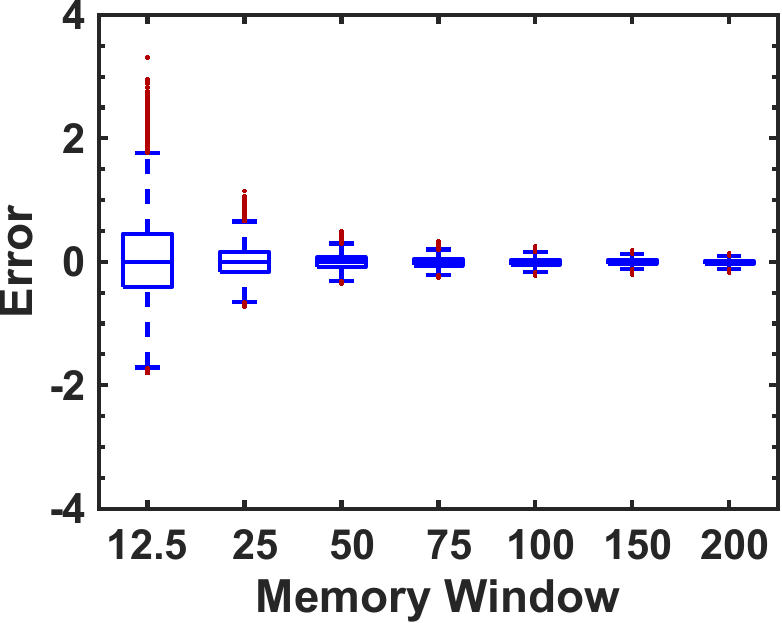}}\\
    \caption{Effect of a) Weight Bits b) Memory Window on VMM error term (w/out non-linearity and C-to-C)}
    \label{fig2}
\vspace{-0mm}
\end{figure}

\section{Methodology}
We propose an end-to-end VMM benchmarking framework called MELISO, which can be described in terms of two distinct stages pertaining to forward and backward computational steps. At the very outset of the forward step, the the matrix and vector inputs are defined in a Python-based environment. Next, device types, crossbar dimensions, and tolerances are specified using a Python module named MelisoPy and a Cython wrapper. The information is then transferred to MLP+NeuroSim, written in C++, to encode the necessary parameters for hardware simulation to compute VMM on RRAM devices. In the backward computational step, the resulting vector of VMM from the forward pass is then scaled and transformed, thus generating the final result as well as performance statistics. Finally, at the output level, the results are collected and analyzed. The entire sequence of steps and the overall design of the MELISO framework is represented in Figure \ref{fig1}.



To perform VMM experiments for this particular study, we considered a population of crossbar arrays with the size of 32 rows and 32 columns. 1000 of $32\times32$ matrices, corresponding to vector \textbf{\textit{A}}, and 1000 of $32\times1$ vectors corresponding to \textbf{\textit{x}} were randomly generated which have then been multiplied using the crossbar to generate 1000 many dot products \textbf{\textit{A.x}} of dimension $32\times1$. The computed values were then compared with the software-calculated dot product to quantify the error. These $32\times1$ error terms were then concatenated creating a $32000\times1$ vector accumulating all the errors from a population of identical devices. This method helps the implementation of a statistically significant number of VMM operations to derive reliable insights into error propagation across devices.

\section{Results}

In Figure \ref{fig2}, we examine the impact of critical device metrics of memory window and weight bit on the error terms observed in VMM operations described in the methodology section. Memory window is defined as the ratio of maximum and minimum conductance levels, $G_{max}/G_{min}$, while weight bit, also referred to as weight precision, corresponds to the maximum number of RRAM conductance states during weight update. We chose \textit{Ag:a-Si} \cite{b22} as a model system due to its reasonable performance in the multi-layer perceptron architecture tested for the MNIST classification task on NeuroSim V3.0 \cite{bx4}. We performed two critical modifications in the default device properties of the \textit{Ag:a-Si} metrics presented in Table \ref{tab1}: i) increased the memory window from the default 12.5 to 100, to accommodate a wider range of conductance states, ii) switched off the C-to-C variation and non-linearity parameters to evaluate the effect of memory window and weight bit metrics independently, without compounding the effects non-linearity and C-to-C variation. These modifications will be rolled back in the subsequent experiments.
\begin{figure}
    \centering
    \subfloat[]{\label{fig3a}\includegraphics[width=.49\linewidth]{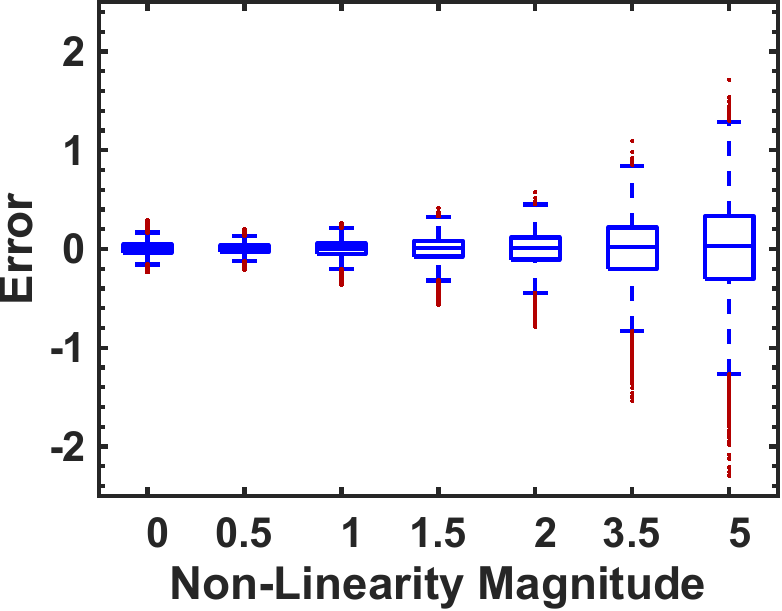}}\hfill
    \subfloat[]{\label{fig3b}\includegraphics[width=.49\linewidth]{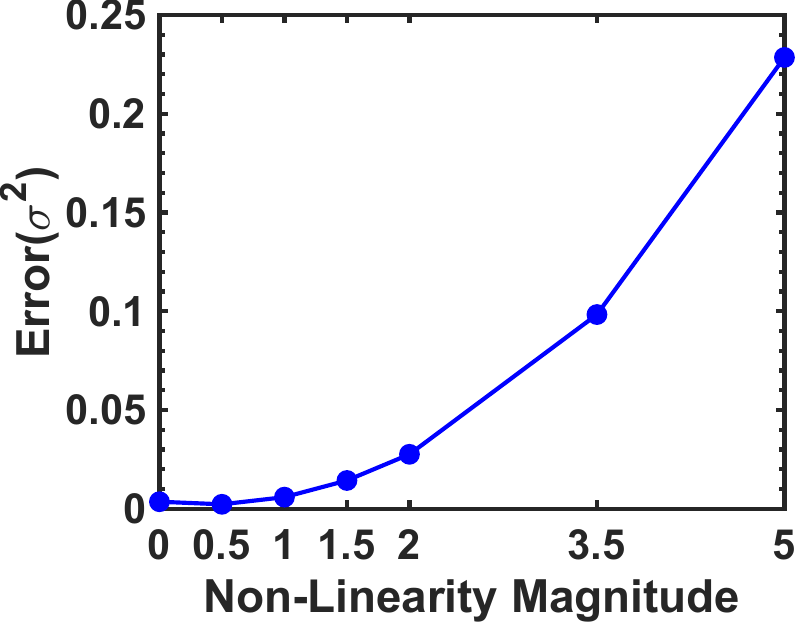}}\\
    \caption{Effect of Non-Linearity on VMM error term}
    \label{fig3}
\end{figure}

Our findings demonstrate a clear dependence of both the magnitude and variance of error terms on the studied device performance metrics. As illustrated in Figure \ref{fig2}a, there is a notable decrease in error magnitude and variance as the weight bits (number of conductance states) increase from 1-bit (2 states) to 11-bit (2048 states). The upper limit of 2048 conductance states was chosen because it represents the current limit in RRAM device technology as the recently reported largest number of conductance states \cite{bx5}. Similarly, Figure \ref{fig2}b shows a reduction in error as the memory window is increased beyond the initial value of 12.5. These findings are consistent with earlier studies, which showed that an increase in the number of weight bits and memory windows would lead to better precision in encoding synaptic weight and reduce the need for weight quantization \cite{bx6,bx7}. This brings our error rates closer to the results of digital computations that use a standard 32-bit floating-point configuration.

\begin{figure*}[!htb]
    \centering
    \subfloat[]{\label{fig4a}\includegraphics[width=0.28\textwidth,keepaspectratio]{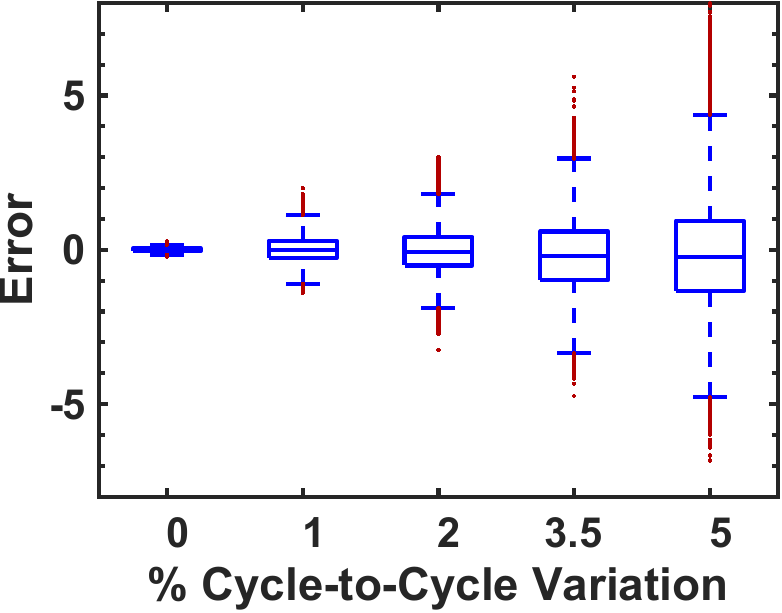}}\hspace{1mm}
    \subfloat[]{\label{fig4b}\includegraphics[width=.28\textwidth,keepaspectratio]{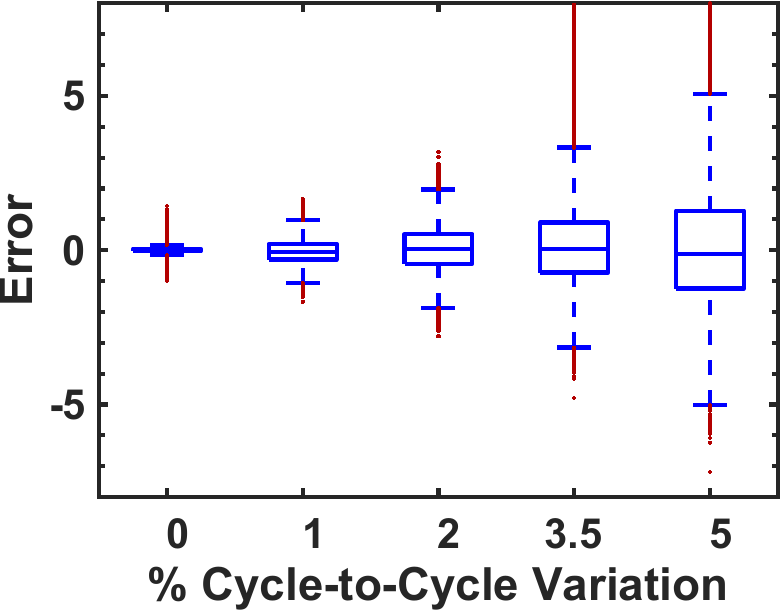}}\hspace{1mm}
    \subfloat[]{\label{fig4c}\includegraphics[width=.28\textwidth,keepaspectratio]{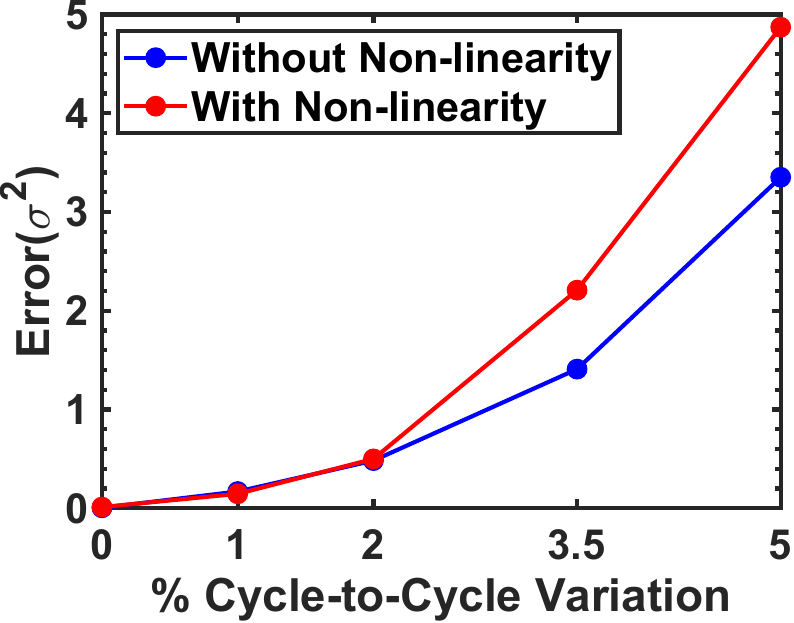}}
    \caption{Effect of the C-to-C variation on VMM error term. a) without considering the non-linearity, b) in the presence of non-linearity, and c) Comparing variance for both cases.}
    \label{fig4}
\end{figure*}


Following our initial examination of primary device metrics, memory window, and weight bits, we further investigated the impact of device non-idealities on performance. Among these non-idealities, weight update non-linearity is particularly critical, as frequently highlighted in the literature \cite{bx6}. It poses a significant challenge to the deployment of RRAM arrays in online training tasks due to its propensity to cause computational errors through incorrect encoding of synaptic weights. This non-linearity renders the implementation of additional algorithms or circuit-level methodologies, such as write-and-verify techniques, essential to mitigate its effects and ensure the reliable operation of RRAM-based systems in real-world applications.

In Figure \ref{fig3}, we explore the effect of weight update non-linearity using the modified model system of \textit{Ag:a-Si}\cite{b22}. This device system has the reported non-linearity metrics, $2.4/$-$4.88$, by default, but we varied the non-linearity magnitude from 0 to 5 in this particular study. Our findings indicate that increases in non-linearity metrics substantially exacerbate the error terms in VMM operations. Additionally, the relationship between error variance and the degree of non-linearity demonstrates an exponential dependency. This dependency reflects the underlying exponential non-linear synaptic weight encoding methodology employed in NeuroSim+\cite{bx3}.


Next, we investigated the impact of C-to-C variation, another significant non-ideality in RRAM devices. C-to-C variation has been demonstrated to be particularly challenging to reduce below certain limits \cite{bx10}. This variability introduces additional errors each time synaptic weights are updated (re-encoded) within the RRAM array. State-of-the-art devices studied in this section (see Table \ref{tab1}) exhibit C-to-C values ranging from 2\% to 5\%. This range represents the performance limitations of the devices in the literature, as C-to-C is often reported as a critical problem. Despite potential mitigations provided by materials, device design, or operational level modifications, RRAMs are inherently more susceptible to C-to-C variations due to the stochastic nature of atomic-level chemical and physical resistive switching mechanisms that are omnipresent in RRAM technology \cite{bx8}.

Figure \ref{fig4} illustrates the relationship between VMM error terms and C-to-C variation in the modified \textit{Ag:a-Si} model system, comparing two configurations with (Figure \ref{fig4a}) and without (Figure \ref{fig4b}) considering non-linearity. Fig. \ref{fig4c} presents the variance comparison of both cases. The data spans a range of C-to-C standard deviations from 0\% to 5\%. We observe a significant increase in the error term as a function of C-to-C, with the highest error rates recorded in this study corresponding to the largest C-to-C values. Even the baseline C-to-C metric of 3.5\% of our model system introduces a substantial level of error, underscoring the critical influence of C-to-C variation on device performance. As expected, the introduction of non-linearity exacerbates the VMM error term, evidenced by the larger variance presented in Figure \ref{fig4c}.

Finally, we conducted a benchmarking study of four different RRAM crossbar systems, with device metrics extracted from the literature (see Table \ref{tab1}) as reported in NeuroSim+ V3.0\cite{bx4}. We evaluated these systems for identical VMM tasks, both with and without considering the effects of device non-idealities such as non-linearity and C-to-C variability. The results of these experiments are depicted in Figure \ref{fig5}, where a and b illustrate the error distributions for scenarios without and with non-idealities, respectively. Additionally, insets in each figure present the VMM error terms as box plots.

Our findings reveal significant differences between the two configurations. In the absence of non-idealities (Figure \ref{fig5a}), the error distributions across the devices are relatively narrow, with the exception of \textit{AlO\textsubscript{x}/HfO\textsubscript{2}}. These devices exhibit similar performance profiles, while \textit{EpiRAM}\cite{b25} stands out with an exceptionally narrow distribution. Conversely, when we accounted for the non-idealities of C-to-C variability and non-linearity (Figure \ref{fig5b}), there was a noticeable increase in both the spread and magnitude of the error distributions. Under these conditions, while the \textit{EpiRAM} still remains the best-performing device, the performance disparity among the other devices becomes more significant.  The \textit{Ag:a-Si
}\cite{b22} and \textit{TaO\textsubscript{x}/HfO\textsubscript{x}}\cite{b23} systems exhibit similar device performances, with the latter presenting an increased 25 and 75 percentile errors. Both systems clearly outperform \textit{AlO\textsubscript{x}/HfO\textsubscript{2}}\cite{b24}.


\begin{figure*}[!htb]
    \centering
    \subfloat[]{\label{fig5a}\includegraphics[width=0.45\textwidth,keepaspectratio]{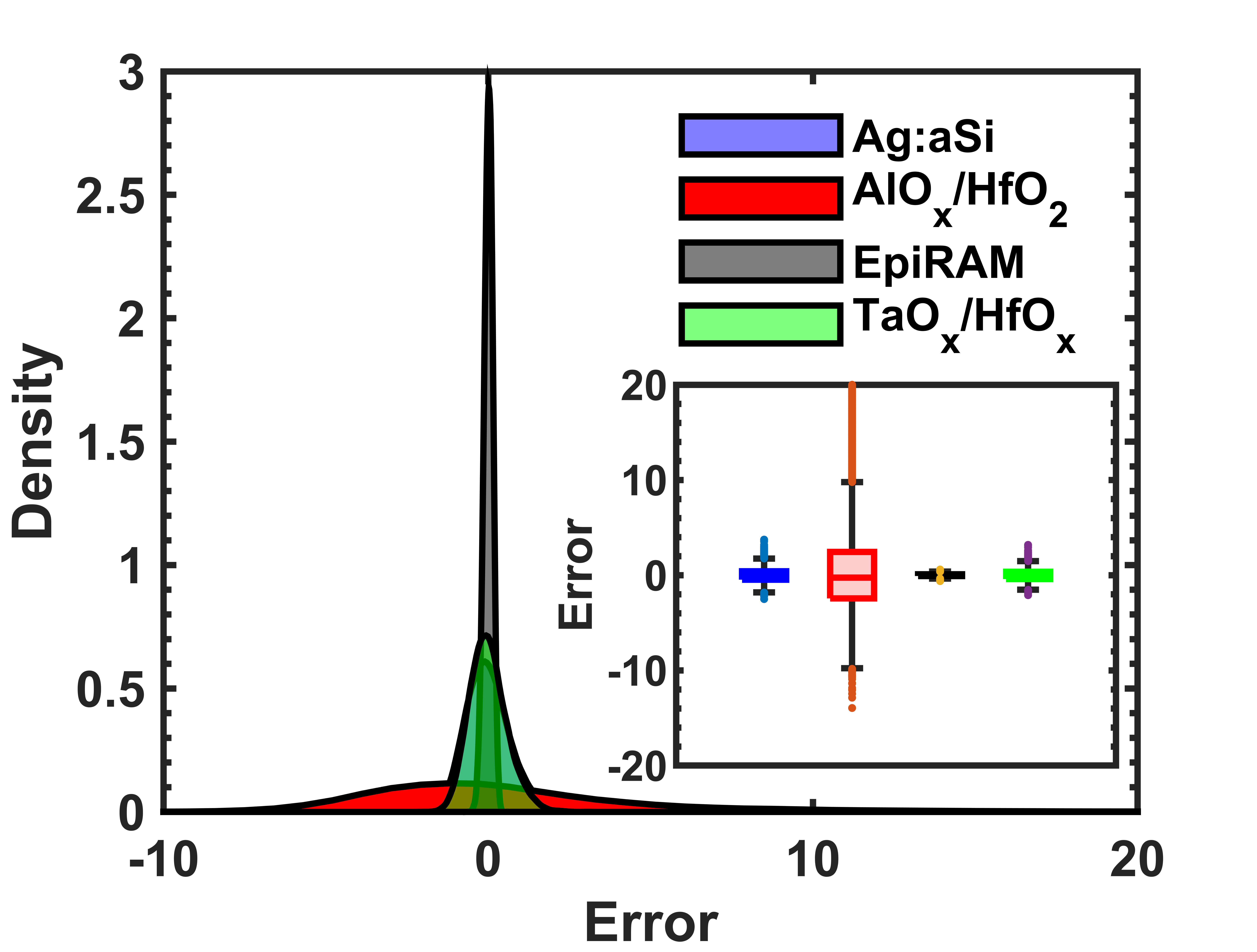}}
    \subfloat[]{\label{fig5b}\includegraphics[width=0.45\textwidth,keepaspectratio]{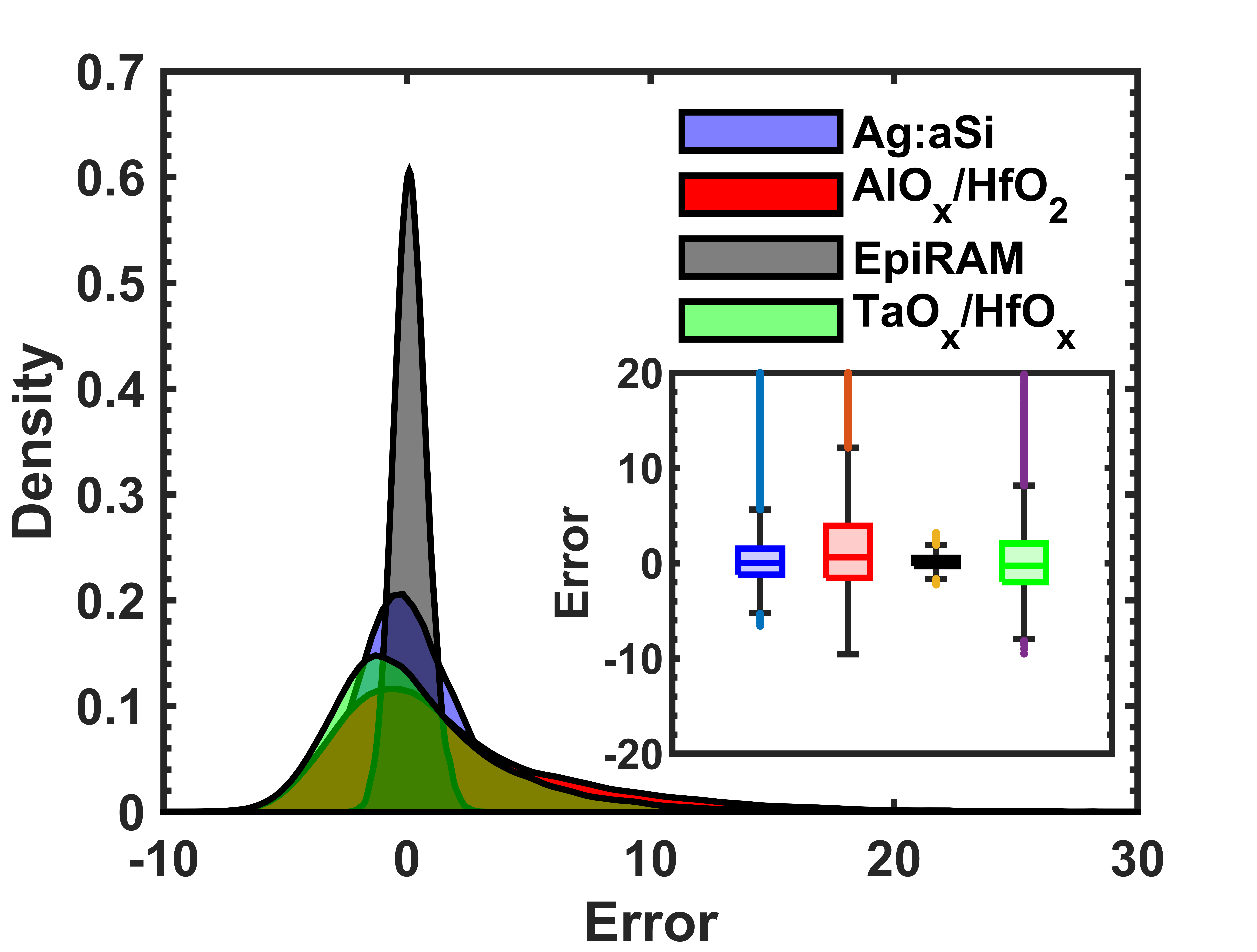}}
    \caption{Effect of non-idealities on VMM performance of different device types, a)Without non-linearity and C-to-C variation, b)With non-linearity and C-to-C variation}
    \label{fig5}
\end{figure*}

These results can be explained by considering the device metrics presented in Table \ref{tab1}. The \textit{EpiRAM} device with the largest memory window (50.2), lowest cumulative weight update non-linearity (0.5/-0.5), and C-to-C variation (2\%) exhibits the smallest error magnitude in VMM operations, thus achieving the highest performance. The second-best performing device, \textit{Ag:a-Si}, demonstrates significant weight update non-linearity (2.4/-4.88) but benefits from a larger memory window of 12.5 and lower C-to-C variation (3.5\%), surpassing both the \textit{TaO\textsubscript{x}/HfO\textsubscript{x}} and \textit{AlO\textsubscript{x}/HfO\textsubscript{2}} systems according to the error box plot in Fig. \ref{fig5b}. Similar error distributions of \textit{Ag:a-Si} and \textit{TaO\textsubscript{x}/HfO\textsubscript{x}} can be attributed to their comparable device metrics: i.e. although \textit{Ag:a-Si} exhibit a higher non-linearity term, memory window, and C-to-C characteristics across these devices are comparable. At the lower end of the performance spectrum, the \textit{AlO\textsubscript{x}/HfO\textsubscript{2}} system exhibits the smallest memory window, coupled with the lowest number of conductance states (weight bit) and C-to-C variation, alongside considerable non-linearity. These device metrics manifest themselves by significantly reducing device performance compared to the other benchmarked devices.

\begin{table*}[]
\caption{Statistical analysis of error distributions for each device material}
\centering
\small
\begin{tabular}{llllllll}
\hline
\multirow{2}{*}{\textbf{Device Type}}        & \multirow{2}{*}{\textbf{Non-linearity}} & \multirow{2}{*}{\textbf{C-to-C}} & \multicolumn{5}{c}{\textbf{Summary Statistics}}                                               \\ \cline{4-8} 
                                             &                                         &                                  & \textbf{Best Fit} & \textbf{Mean} & \textbf{Variance} & \textbf{Skewness} & \textbf{Kurtosis} \\ \hline
\multicolumn{1}{c}{\multirow{2}{*}{\hspace{-0.55cm}Ag:a-Si}} & No                                      & No                               & Normal-3-Mixture  & -0.00084      & 0.4607            & 0.4639            & 0.4369            \\
\multicolumn{1}{c}{}                         & Yes                                     & Yes                              & Johnson $S_u$        & 0.7059        & 13.0763           & 3.3405            & 15.6567           \\ \hline
\multirow{2}{*}{AlO\textsubscript{x}/HfO\textsubscript{2}}               & No                                      & No                               & Normal-3-Mixture  & 0.8311        & 32.0761           & 2.7935            & 13.3362           \\
                                             & Yes                                     & Yes                              & Normal-3-Mixture  & 0.5247        & 13.9694           & 1.5065            & 3.7796            \\ \hline
\multirow{2}{*}{EpiRAM}                      & No                                      & No                               & SHASH             & 0.0044        & 0.0179            & -0.2463           & 0.0256            \\
                                             & Yes                                     & Yes                              & Normal-2-Mixture  & 0.1453        & 0.4630            & 0.1927            & 0.1744            \\ \hline
\multirow{2}{*}{TaO\textsubscript{x}/HfO\textsubscript{x}}                   & No                                      & No                               & Normal-3-Mixture  & -0.0001       & 0.3336            & 0.4314            & 0.3761            \\
                                             & Yes                                     & Yes                              & Normal-3-Mixture  & 0.4117        & 12.5167           & 1.2150            & 2.2775            \\ \hline
\end{tabular}
\label{tab2}
\end{table*}

It should be noted that the performance of \textit{EpiRAM}, \textit{Ag:a-Si} and \textit{TaO\textsubscript{x}/HfO\textsubscript{x}} deteriorates with the introduction of non-linearity and C-to-C variation. These observations are consistent with our earlier findings illustrated in Figure \ref{fig2}, \ref{fig3}, and \ref{fig4}, where we systematically explored the impact of memory window, non-linearity, and C-to-C variability on error dynamics. An interesting pattern emerges when we closely study the performance trends of \textit{AlO\textsubscript{x}/HfO\textsubscript{2}}. Due to low conductance states, poor memory window, and high C-to-C variation, the variability introduced through device non-idealities slightly improves device performance. This is specifically prominent when we study the outliers: i.e. as shown in Figure \ref{fig5}, the $25$ and $75$ percentile errors increase slightly while shifting from ideal to non-ideal \textit{AlO\textsubscript{x}/HfO\textsubscript{2}} performances, however, the noise introduced by nonidealities reduces the span of outliers, resulting in a slight improvement to distributional metrics of the errors, as shown in Table \ref{tab2}.


Table \ref{tab2} presents an overview of the empirical distribution of the errors with best fitting models and the corresponding distribution moment characteristics for \textit{Ag:a-Si}, \textit{TaO\textsubscript{x}/HfO\textsubscript{x}}, \textit{AlO\textsubscript{x}/HfO\textsubscript{2}} and \textit{EpiRAM}, across two key device non-ideality parameters: e.g. non-linearity and C-to-C. Our results indicate that the errors typically follow one of the following distributions: Johnson $S_u$, Normal-3-Mixture, Normal-2-Mixture, and Sinh-Arc-Sinh (SHASH). The error distributions do not follow a typical normal distribution due to the presence of asymmetry and heavy tail behavior, which are measured through skewness and kurtosis metrics, respectively. The analysis shows that among the four devices studied in this paper, \textit{AlO\textsubscript{x}/HfO\textsubscript{2}} exhibits the highest variance, which is followed by \textit{Ag:a-Si} and \textit{TaO\textsubscript{x}/HfO\textsubscript{x}}. This trend is seen in our experiments with both ideal and non-ideal device parameters.

Studying the skewness and kurtosis allows us to understand trends in error distribution that cannot be explained through box plots or mean and variance metrics alone. As indicated in Table \ref{tab2}, \textit{Ag:a-Si} (with non-idealities) and \textit{AlO\textsubscript{x}/HfO\textsubscript{2}} (without non-idealities) exhibit high positive skewness indicating a longer tail on the right side of the distribution. \textit{Ag:a-Si} with non-linearity and C-to-C variability also has the highest kurtosis indicating a heavier tail compared to \textit{EpiRAM} or \textit{TaO\textsubscript{x}/HfO\textsubscript{x}}. This analysis suggests that the skewness and kurtosis metrics are most sensitive to the device non-linearity properties: Although \textit{Ag:a-Si} has better memory window and C-to-C properties compared to \textit{TaO\textsubscript{x}/HfO\textsubscript{x}}, high non-linearity in \textit{Ag:a-Si} causes its skewness and kurtosis metrics to be higher. More specifically, we observe that the non-linearity impacts are becoming more prominent only after going into the third and fourth moments (e.g. skewness and kurtosis) of the error distributions. 





\section{Conclusion and Outlook}

Successful deployment of RRAM devices for in-memory computing applications relies on conducting accurate VMM operations with improved energy efficiency and reduced latency metrics. This study introduces a comprehensive benchmarking framework, MELISO, designed to methodically analyze and model the error landscape of VMM operations conducted on RRAM crossbar systems, with the ultimate goal of guiding the development of more robust RRAM technologies and algorithms. Our extensive benchmarking reveals significant insights into the impact of RRAM device parameters and non-idealities on error propagation. Key findings include distinct patterns in error distributions related to specific device characteristics, such as C-to-C variability and non-linear responses. Furthermore, we have reported the parametric distributions that most accurately represent the observed errors, providing a solid foundation for future algorithmic strategies aimed at either mitigating or harnessing these errors to enhance computational accuracy in memory-centric computing environments.

Future research will focus on enhancing our understanding of error propagation dynamics and their computational implications. Specifically, we plan to investigate the aspects of neuromorphic device virtualization and parallelization primitives for RRAM arrays. Additionally, our future work also involves the development of a suite of computationally efficient, general-purpose optimization libraries for convex and non-convex solvers, to address low-latency problems in power systems engineering.  We will also conduct a more detailed study on optimizing RRAM device metrics for targeted computational tasks while considering performance and energy consumption benchmarking metrics. Through such concerted efforts, RRAM-based systems are expected to realize their full potential, contributing substantially to the next generation of computational technology.

\section{Acknowledgment}
Gozde Tutuncuoglu would like to acknowledge NSF-CRII grant 2153177.

\vspace{12pt}

\end{document}